\documentclass{elsart5p}

\usepackage{graphicx}
\usepackage{amssymb}
\usepackage{natbib}

\begin{document}

\begin{frontmatter}

\title{Probing non-Riemannian spacetime geometry}

\author{Dirk Puetzfeld\corauthref{cor1}}
\address{Institute of Theoretical Astrophysics, University of Oslo, P.O.\ Box 1029, 0315 Oslo, Norway}
\corauth[cor1]{Also at: Max-Planck-Institute for Gravitational Physics (Albert-Einstein-Institute), Am Muehlenberg 1, 14476 Golm, Germany}
\ead{dirk.puetzfeld@astro.uio.no}
\ead[url]{http://www.thp.uni-koeln.de/$\sim$dp}

\author{Yuri N. Obukhov\corauthref{cor2}}
\corauth[cor2]{Also at: Department of Theoretical Physics, Moscow State University, 117234 Moscow, Russia}
\address{Institute for Theoretical Physics, University of Cologne, Z\"ulpicher Stra\ss e 77, 50937 K\"oln, Germany}
\ead{yo@thp.uni-koeln.de}

\begin{abstract}
The equations of motion for matter in non-Riemannian spacetimes are derived via a multipole method. It is found that only test bodies with microstructure couple to the non-Riemannian spacetime geometry. Consequently it is impossible to detect spacetime torsion with the satellite experiment Gravity Probe B, contrary to some recent claims in the literature.
\end{abstract}

\end{frontmatter}

\section{Introduction.}\label{introduction_sec}

How do test particles move under the influence of the gravitational field? In the context of the theory of General Relativity (GR) this question was attacked nearly over seventy years ago \cite{Mathisson:1931:3,Mathisson:1931:1,Mathisson:1931:2}. Since then the relation between the field equations and the equations of motion within gravitational theories has been subject to many investigations \cite{Robertson:1937,Fock:1939,Mathisson:1937,Papapetrou:1940:1,InfeldSchild:1949,Papapetrou:1951:3,Papapetrou:1951,Tulczyjew:1959,HavasGoldberg:1962}. The intimate link between these equations is the feature of General Relativity which distinguishes it from other physical theories. 

As it is well known, the Riemannian geometry of spacetime can be tested with structureless particles (with or without rest mass). An interesting physical question is whether the latter can also probe more general non-Riemannian geometries that possibly could arise on a spacetime manifold. There are claims in the literature that the answer is positive. In \cite{Mao:etal:2006}, for example, the detectability of the spacetime torsion is discussed in the context of the satellite experiment Gravity Probe B.  

In this letter, we demonstrate that structureless particles can only test the Riemannian geometry, and that they are not affected by the non-Riemannian geometrical structures of spacetime. In order to prove this, we systematically derive the equations of motion of matter in the metric-affine gravity (MAG) theory \cite{Hehl:1995}, which provides a proper physical and mathematical framework for gravitational models with non-Riemannian structures of spacetime. We thereby confirm and extend earlier results in the context of Riemann-Cartan geometries \cite{Hehl:1971,Trautman:1972,Stoeger:Yasskin:1979,Stoeger:Yasskin:1980,
Nomura:Shirafuji:Hayashi:1991}, for which it was shown that only the intrinsic spin of test matter couples to spacetime torsion.
Note that we consider macroscopic classical matter in this letter. The analysis of
the dynamics of quantum particles with spin in the Riemann-Cartan spacetime
can be found in \cite{Audretsch:1981:1,Bragov:etal:1991:1,Bragov:etal:1991:2,Nomura:Shirafuji:Hayashi:1992,Shapiro:2002}.

In GR, the mass, or more precisely the energy-momentum of matter is the only physical source of the gravitational field. The energy-momentum current corresponds to the local translational (or the diffeomorphism) spacetime symmetry. In MAG, this symmetry is extended to the local affine group that is a semi-direct product of translations times the local linear spacetime symmetry group. Via the Noether theorem, such a symmetry gives rise to additional conserved currents that describe microscopic characteristics of matter. In continuum mechanics \cite{Cosserat:1909,Weyssenhoff:Raabe:1947,Kroener:1958,Truesdell:Toupin:1960,Mindlin:1964,Capriz:1989}, such matter is known as a medium with microstructure. In physical terms, this means that the elements of a material continuum have internal degrees of freedom: spin, dilation and shear. These three microscopic sources are irreducible parts (that correspond, respectively, to the Lorentz, dilational and shear-deformational subgroups of the general linear group) of the hypermomentum current. 

The geometry that arises on the spacetime manifold is non-Riemannian, with nontrivial curvature, torsion, and nonmetricity. The resulting general scheme of MAG embeds a wide spectrum of gauge gravitational models based on the Poincar\'e, conformal, Weyl, de Sitter, and other spacetime symmetry groups (for an overview, see \cite{Hehl:1995}, for example). 

The energy-momentum current and the hypermomentum current (spin + dilaton + shear current) are the sources of the gravitational field in MAG. Accordingly, test bodies, that are formed of matter with microstructure, have two kinds of physical properties which determine their dynamics in a curved spacetime. The properties of the first type have {\it microscopic} origin, they arise directly from the fact that the elements of a medium have internal degrees of freedom (microstructure). The properties of the second type are essentially {\it macroscopic}, they arise from the collective dynamics of matter elements characterized by mass (energy) and momentum. Hence, the qualitative picture is as follows: The averaging of the microscopic hypermomentum current yields the integrated spin, dilaton, and shear charge of a test body. In addition, the averaging of the energy-momentum and of its multipole moments gives rise to the {\it orbital} integrated momenta. The well known first moment is the  orbital angular momentum. It describes the behavior of a test particle as a rigid body, that is, its rotation. In addition, there are first orbital moments that describe {\it deformations} of a body. These are the orbital dilation momentum (that describes isotropic volume expansion) and the orbital shear momentum (that determines the anisotropic deformations with fixed volume). The three together (orbital angular momentum, orbital dilation momentum, and orbital shear momentum) comprise the generalized integrated orbital momentum. In this letter, we compare the gravitational interaction of the integrated hypermomentum to that of the integrated orbital momentum of a {\it rotating and deformable test body}. Thereby, we generalize the previous analysis \cite{Stoeger:Yasskin:1980} in which the effects of the integrated spin were compared to the effects of the orbital angular momentum of a rotating rigid test body. 
 
\section{Metric-affine gravity.}\label{mag_intro_sec}

For a review of the MAG theory see \cite{Hehl:1995,Gronwald:Hehl:1996}, and references therein. In this theory, besides the usual ``weak'' long-range Newton-Einstein type gravity, described by the metric $g_{ij}$ of spacetime, an additional ``strong'' short-range gravity piece is mediated by the independent linear connection $\Gamma_{i j}{}^{k}$. It is different from the Riemannian (Christoffel) connection, and the difference is described in terms of the tensors of nonmetricity $Q_{ijk}:=-\nabla_i g_{j k}$ and of the torsion $S_{ij}{}^{k}:=\Gamma_{i j}{}^{k}-\Gamma_{j i}{}^{k}$ which are also manifest in the non-Riemannian pieces of the curvature $R_{ijk}{}^{l}$.

The matter currents, which are the sources of the gravitational field, are obtained by variation of the matter Lagrangian with respect to the gravitational potentials (metric $g_{ij}$, coframe $h^\alpha_j$, connection $\Gamma_{i j}{}^{k}$). This yields the canonical energy-momentum $T_i{}^j:=h^\alpha_i\delta {L}_{\rm mat} / \delta h^\alpha_j$, the metrical energy-momentum $t^{ij}:=2 \delta {L}_{\rm mat} / \delta g_{ij}$, and the hypermomentum $\Delta^i{}_j{}^k := \delta { L}_{\rm mat} / \delta\Gamma_{ki}{}^j$ current.
 
The conservation laws of the theory, cf.\ \cite{Obukhov:Rubilar:2006} for a recent review, serve as starting point for the derivation of the propagation equations for the multipole moments of the matter currents. 

\section{Energy-momentum conservation.}\label{em_conservation_subsec}

The Noether theorem for the diffeomorphism invariance of the matter action yields the conservation law of the energy-momentum 
\begin{equation}
{\stackrel{\{\,\}}{\nabla}}_j\left(T_i{}^j - N_{ikl}\,\Delta^{klj}\right)=\big({\stackrel{\{\,\}}{R}}_{ijkl} - {\stackrel{\{\,\}}{\nabla}}_i \,N_{jkl}\big)\Delta^{klj}.\label{DT1}
\end{equation}
Here, and in the following, curled braces ``$\{ \}$'' denote objects which are based on the symmetric Riemannian connection (Christoffel symbols), and $N_{ij}{}^k :={\stackrel{\{\,\}}{\Gamma}}_{ij}{}^k-\Gamma_{ij}{}^k$ represents the so-called distorsion tensor. Equation (\ref{DT1}) can be identically rewritten as 
\begin{equation}
{\stackrel{\{\,\}}{\nabla}}_j\,T_i{}^j = \widehat{R}_{ijkl}\,\Delta^{klj} + N_{ikl}\,{\stackrel{\{\,\}}{\nabla}}_j\Delta^{klj},\label{DT2}
\end{equation}
where we introduced $\widehat{R}_{ijkl} := {\stackrel{\{\,\}}{R}}_{ijkl} - {\stackrel{\{\,\}}{\nabla}}_i N_{jkl} + {\stackrel{\{\,\}}{\nabla}}_j N_{ikl}$.

\section{Hypermomentum conservation.}\label{hm_conservation_subsec}

The Noether theorem for the invariance of MAG under the local general linear group yields (on the ``mass shell", i.e., when the matter satisfies the field equations): 
\begin{equation}
{\stackrel{\{\,\}}{\nabla}}_j\,\Delta^{klj} - N_{ij}{}^k\Delta^{jli} + N^{jli}\Delta^k{}_{ij} + T^{lk} - t^{kl} = 0.\label{GLc}
\end{equation}

\section{Propagation equations.}\label{sec_propa_eq_ys}

Denoting the densities of objects by a tilde ``$\widetilde{{\phantom{A}}}$'' the conservation equations for the canonical energy-momentum current (\ref{DT2}) and hypermomentum current (\ref{GLc}), take the following form  
\begin{eqnarray}
\partial_j\widetilde{T}{}_{i}{}^{j} &=&R_{ijk}{}^{l} \widetilde{\Delta }^{k}{}_{l}{}^{j}+\Gamma _{ij}{}^{k} \widetilde{T}{}_{k}{}^{j}+N_{ij}{}^{k}\widetilde{t}^{j}{}_{k}, \label{em_conservation_ala_DY_1} \\
\partial_j\widetilde{\Delta }^{k}{}_{l}{}^{j} &=&\Gamma _{jl}{}^{m} \widetilde{\Delta }^{k}{}_{m}{}^{j}-\Gamma _{mj}{}^{k} \widetilde{\Delta }^{j}{}_{l}{}^{m}-\widetilde{T}{}_{l}{}^{k}+\widetilde{t}^{k}{}_{l}. \label{em_conservation_ala_DY_2}
\end{eqnarray}
Note that $\Gamma_{ij}{}^{k}$ represents the full connection, the last two equations should be compared to (42) and (43) in \cite{Stoeger:Yasskin:1980}. 

\section{Conservation equations integrated.}

We introduce the integrated multipole moments as follows:
\begin{eqnarray}
\underline{\Delta}^{b_{1}\cdots b_{n}i}{}_{j}{}^{k} &:&=\int \left( \prod\limits_{\alpha=1}^{n}\delta x^{b_{\alpha }}\right) \widetilde{\Delta }^{i}{}_{j}{}^{k}, \nonumber \\
\underline{T}^{b_{1}\cdots b_{n}}{}_{i}{}^{j} &:&=\int \left( \prod\limits_{\alpha =1}^{n}\delta x^{b_{\alpha }}\right) \widetilde{T}_{i}{}^{j},  \nonumber \\
\underline{t}^{b_{1}\cdots b_{n}i}{}_{j} &:&=\int \left( \prod\limits_{\alpha=1}^{n}\delta x^{b_{\alpha }}\right) \widetilde{t}^{i}{}_{j}. \label{DY_int_moments_definitions}
\end{eqnarray}
The integrals are taken over a 3-dimensional slice $\Sigma(t)$, at a time $t$,
of the world tube of a test body. We use the condensed notation
\begin{equation}
\int\,f = \int_{\Sigma(t)}\,f(x)\,d^3x.
\end{equation} 
With these definitions the integrated conservation laws (\ref{em_conservation_ala_DY_1}) and (\ref{em_conservation_ala_DY_2}) take the following form (an inverted circumflex, e.g.\ ``$\check{b}_\beta$'', indicates the omission of an index from a list and $v^a:=dY^a/dt$, cf. Fig.~\ref{fig_world_tube})

\begin{eqnarray}
\hspace{-2mm} &&\frac{d}{dt}\underline{T}^{b_{1}\cdots b_{n}}{}_{i}{}^{0} =\sum_{\beta =1}^{n}\left(\underline{T}^{b_{1}\cdots \check{b}_{\beta } \cdots b_{n}}{}_{i}{}^{b_{\beta}}-v^{b_{\beta }}{}\underline{T}^{b_{1}\cdots \check{b}_{\beta }\cdots b_{n}}{}_{i}{}^{0}\right) \nonumber \\
\hspace{-2mm} &&+\int \left( \prod \limits_{\alpha =1}^{n}\delta x^{b_{\alpha }}\right) \left( R_{ijk}{}^{l} \widetilde{\Delta }^{k}{}_{l}{}^{j}+\Gamma _{ij}{}^{k}  \widetilde{T}{}_{k}{}^{j}+N_{ij}{}^{k}\widetilde{t}^{j}{}_{k}\right), \label{DY_int_cons_eq_1} \\
\hspace{-2mm} &&\frac{d}{dt}\underline{\Delta}^{b_{1}\cdots b_{n}k}{}_{l}{}^{0} =\sum_{\beta =1}^{n}\left(\underline{\Delta}^{b_{1}\cdots \check{b}_{\beta } \cdots b_{n}k}{}_{l}{}^{b_{\beta}}-v^{b_{\beta }}{}\underline{\Delta}^{b_{1} \cdots \check{b}_{\beta }\cdots b_{n}k}{}_{l}{}^{0}\right)\nonumber \\
\hspace{-2mm} &&+\int \left( \prod \limits_{\alpha =1}^{n}\delta x^{b_{\alpha }}\right) \left( \Gamma_{jl}{}^{m} \widetilde{\Delta }^{k}{}_{m}{}^{j}-\Gamma_{mj}{}^{k} \widetilde{\Delta}^{j}{}_{l}{}^{m}-\widetilde{T}{}_{l}{}^{k}+\widetilde{t}^{k}{}_{l}\right).  \label{DY_int_cons_eq_2}
\end{eqnarray}

Equations (\ref{DY_int_cons_eq_1}) and (\ref{DY_int_cons_eq_2}) should be compared to equations (51) and (52) in \cite{Stoeger:Yasskin:1980}.

\section{Propagation equations for pole-dipole particles.}

\begin{figure}
\begin{center}
\includegraphics[width=8cm]{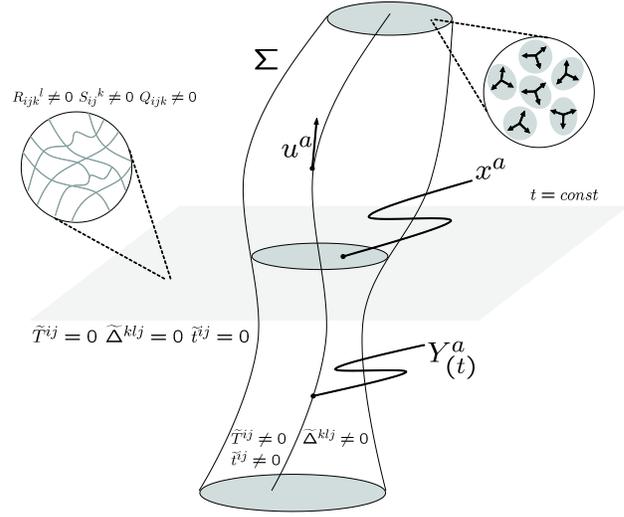}
\end{center}
\caption{\label{fig_world_tube} Sketch of the hypersurface $\Sigma$, i.e., the world tube of the test particle. A continuous curve through the tube is parametrized by $Y^a$. Coordinates within the world tube with respect to a coordinate system centered on $Y^a$ are labeled by $x^a$. The velocity along the world line is denoted by $u^a:=dY^a/ds$ with $u^0=dt/ds$.}
\end{figure}

{}From the general expressions (\ref{DY_int_cons_eq_1}) and (\ref{DY_int_cons_eq_2}) we can derive the propagation equations for pole-dipole particles. For such bodies the following moments are non-vanishing: $\underline{\Delta}^{i}{}_{j}{}^{k}, \underline{T}_{i}{}^{j}, \underline{T}^{i}{}_{j}{}^{k}, \underline{t}^{i}{}_{j},$ and $\underline{t}^{ij}{}_{k}$. The expansion of geometrical quantities around the worldline $Y(t)$ of the body, cf. Fig.~\ref{fig_world_tube}, into a power series in $\delta x^{a}=x^{a}-Y^{a},$ reads
\begin{eqnarray}
\left. R_{ijk}{}^{l} \right| _{x} &=&\left. R_{ijk}{}^{l} \right| _{Y} +\delta x^{a}\left. R_{ijk}{}^{l}{}_{,a}\right| _{Y}+\dots ,  \nonumber \\
\left. \Gamma _{ij}{}^{k}\right| _{x} &=&\left. \Gamma _{ij}{}^{k}\right| _{Y}+\delta x^{a}\left. \Gamma _{ij}{}^{k}{}_{,a}\right| _{Y}+\dots , \nonumber \\
\left. N_{ij}{}^{k}\right| _{x} &=&\left. N_{ij}{}^{k}\right| _{Y}+\delta x^{a} \left. N_{ij}{}^{k}{}_{,a}\right| _{Y}+\dots . \label{DY_geom_quant_taylor}
\end{eqnarray}
The general form of the integrated conservation laws (\ref{DY_int_cons_eq_1}) and (\ref{DY_int_cons_eq_2}) then yields the following set of propagation equations:
\begin{eqnarray}
\frac{d}{dt} \underline{T}_{i}{}^{0} &=&R_{ijk}{}^{l} \underline{\Delta}^{k}{}_{l}{}^{j}+\Gamma _{ij}{}^{k}\underline{T}_{k}{}^{j} +\Gamma _{ij}{}^{k}{}_{,a} \underline{T}^{a}{}_{k}{}^{j} \nonumber \\
&&+N_{ij}{}^{k} \underline{t}^{j}{}_{k}+N_{ij}\,^{k}{}_{,a} \underline{t}^{a}{}^{j}{}_{k},  \label{DY_pd_prop_eq_1} \\
\frac{d}{dt}\underline{T}^{a}{}_{i}{}^{0} &=&\underline{T}_{i}{}^{a}-v^{a} \underline{T}_{i}{}^{0}+\Gamma _{ij}{}^{k}\underline{T}^{a}{}_{k}{}^{j} +N_{ij}{}^{k} \underline{t}^{a}{}^{j}{}_{k},  \label{DY_pd_prop_eq_2} \\
0 &=& \underline{T}^{b}{}_{i}{}^{a}+\underline{T}^{a}{}_{i}{}^{b}-v^{a} \underline{T}^{b}{}_{i}{}^{0}-v^{b}\underline{T}^{a}{}_{i}{}^{0}, \label{DY_pd_prop_eq_3} \\
\frac{d}{dt}\underline{\Delta}^{k}{}_{l}{}^{0} &=&\Gamma _{jl}{}^{m} \underline{\Delta}^{k}{}_{m}{}^{j}-\Gamma _{mj}{}^{k} \underline{\Delta}^{j}{}_{l}{}^{m}-\underline{T}_{l}{}^{k}+\underline{t}^{k}{}_{l},  \label{DY_pd_prop_eq_4} \\
0 &=&\underline{\Delta}^{k}{}_{l}{}^{a}-v^{a}\underline{\Delta}^{k}{}_{l}{}^{0} -\underline{T}^{a}{}_{l}{}^{k}+\underline{t}^{ak}{}_{l}.  \label{DY_pd_prop_eq_5}
\end{eqnarray}
Here we suppressed the dependencies on the points at which certain quantities are evaluated. The set (\ref{DY_pd_prop_eq_1})-(\ref{DY_pd_prop_eq_5}) represents the generalization of the propagation equations for pole-dipole particles to metric-affine gravity. 

\section{Analyzing the propagation equations.}

Before we study the propagation of massive bodies in the gravitational field, it is worthwhile to recall some well known facts about the dynamics of the electrically charged bodies in the electromagnetic field. The electric 4-current density $\widetilde{J}^i$ is the primary object then, with $\widetilde{\rho} =\widetilde{J}^0$ the electric charge density. When the size of the body is much smaller than the typical length over which the fields change significantly, it can be treated as a test particle. The motion of the latter is conveniently described by the interaction of the particle's multipole moments $\underline{J}{}^{b_1\cdots b_n k} = \int\delta x^{b_1}\cdots\delta x^{b_n}\widetilde{J}^k$ with the electric and magnetic fields. Normally, the lowest moments are most important and they sufficiently well determine the behavior of the body. In particular, the zeroth moment $Q = \underline{J}^0 = \int \widetilde{\rho}$ is just the total electric charge of the body, the first moment $D^i = \underline{J}^{i0} = \int \delta x^i\widetilde{\rho}$ is the electric dipole, and so on. 

We proceed along the same lines for the dynamics of gravitating particles by replacing the electromagnetic field with the gravitational one, and the electric current with the energy-momentum and hypermomentum currents. Then, we naturally define the integrated quantities as follows: $\underline{P}_i := \underline{T}_i{}^0$ is the total 4-momentum of the body (recall that $p_i =\widetilde{T}_i{}^0$ is the density of the energy $\widetilde{T}_0{}^0$ and momentum $\widetilde{T}_a{}^0$, $a=1,2,3$, of matter), $\underline{L}^k{}_l := \underline{T}^k{}_l{}^0 = \int \delta x^k p_l$ the total orbital canonical energy-momentum. The antisymmetrized (over the indices $k$ and $l$) quantity is the most familiar orbital momentum which naturally arises for {\it rigid} bodies. However, since we study the general case of {\it deformable} bodies, the symmetric part of the first moment is now relevant too. Furthermore, we introduce $\underline{Y}^k{}_l := \underline{\Delta}^k{}_l{}^0$ as the integrated intrinsic hypermomentum, and define
\begin{equation}
{\cal P}_{i}:=\underline{P}_{i}-N_{ik}{}^{l}\underline{Y}^{k}{}_{l} -{\stackrel{\{\,\}}{\Gamma}}_{ik}{}^{l}\underline{L}^{k}{}_{l}{}, \label{Ptot}
\end{equation}
the generalized total 4-momentum. Albeit this definition appears to be ``natural'' in the context of MAG -- and actually prolongs the one known from \cite{Stoeger:Yasskin:1980} -- one should be clear about the fact that it does not necessarily correspond to a directly measurable quantity. In addition, we introduce a shorter notation for the ``convective currents": For the intrinsic hypermomentum we have ${\stackrel{(c)}{\underline{\Delta}}}{}^k{}_l{}^m := \underline{\Delta}^k{}_l{}^m - v^m\,\underline{\Delta}^k{}_l{}^0$, and for the orbital canonical energy-momentum ${\stackrel {(c)}{\underline{T}}}{}^k{}_l{}^m :=\underline{T}^k{}_l{}^m - v^m\,\underline{T}^k{}_l{}^0$. The fluid derivative is defined as follows $\nabla_v\,\underline{Y}^i{}_k:=d/dt \, \underline{Y}^i{}_k + v^m \Gamma_{mj}{}^{i} \underline{Y}^j{}_k - v^m \Gamma_{mk}{}^{j} \underline{Y}^i{}_j$. With this notation, we recast the propagation equations (\ref{DY_pd_prop_eq_1})-(\ref{DY_pd_prop_eq_5}) into 
\begin{eqnarray}
 {\stackrel{\{\,\}}{\nabla}}_{v}{\cal P}_{i}&=& \left({\stackrel{\{\,\}}{R}}_{ijk}{}^{l}-{\stackrel{\{\,\}}{\nabla}}_{i} N_{jk}{}^{l}\right)\underline{\Delta}^{k}{}_{l}{}^{j}+{\stackrel{\{\,\}}{R}}_{ijk}{}^{l} {\stackrel{(c)}{\underline{T}}}{}^{k}{}_{l}{}^{j} \nonumber \\
&&+ {\stackrel{\{\,\}}{R}}_{kji}{}^{l}\,\underline{L}^k{}_{l}\,v^j,\label{DY_pd_prop_eq_1a} \\
 \underline{T}_k{}^i &=& v^i\,\underline{P}_k + {\frac {d}{dt}} \,\underline{L}^i{}_k - {\stackrel{\{\,\}}{\Gamma}}_{kj}{}^{l} \, \underline{T}^i{}_l{}^j + N_{kj}{}^{l}\,{\stackrel {(c)}{\underline{\Delta}}}{}^j{}_l{}^i,\label{DY_pd_prop_eq_2a}\\
 {\stackrel {(c)}{\underline{T}}}{}^{(a}{}_i{}^{b)} &=& 0, \label{DY_pd_prop_eq_3a}\\
 \nabla_v\,\underline{Y}^i{}_k &=& -\,\underline{T}_k{}^i + \underline{t}^i{}_k - \Gamma_{jl}{}^{i}\,{\stackrel {(c)}{\underline{\Delta}}} {}^l{}_k{}^j +\Gamma_{jk}{}^{l}\,{\stackrel {(c)}{\underline{\Delta}}}{}^i {}_l{}^j, \label{DY_pd_prop_eq_4a}\\
 {\stackrel {(c)}{\underline{\Delta}}}{}^k{}_l{}^a &=& \underline{T}^a{}_l{}^k - \underline{t}^{ak}{}_l.\label{DY_pd_prop_eq_5a}
\end{eqnarray}
The propagation equation (\ref{DY_pd_prop_eq_1a}) for the generalized total 4-momentum should be compared to (53) in \cite{Stoeger:Yasskin:1980}. Equation (\ref{DY_pd_prop_eq_2a}) describes the canonical energy-momentum in terms of the usual combination of the ``translational" plus ``orbital" contributions (the first two terms), plus the additional contribution of the first moments. Equation (\ref{DY_pd_prop_eq_3a}) simply tells us that the convective current ${\stackrel{(c)}{\underline{T}}}{}^{a}{}_i{}^{b}$ is antisymmetric in the upper indices $a$ and $b$. The next equation (\ref{DY_pd_prop_eq_4a}) is actually an equation of motion for the intrinsic hypermomentum. Its  form closely follows the Noether conservation law of the hypermomentum, cf.\ (\ref{GLc}). Finally, the equation (\ref{DY_pd_prop_eq_5a}) expresses the convective intrinsic hypermomentum current in terms of the first moments of the energy-momentum. 

\section{Physical consequences.}

From the set (\ref{DY_pd_prop_eq_1a})-(\ref{DY_pd_prop_eq_5a}) we notice a general feature that characterizes the coupling between the physical objects (currents) with the geometrical objects (metric, connection, and the derived quantities). Namely, the {\it intrinsic} current (the one that is truly {\it microscopic}, which arises from the averaging over the medium with the elements with microstructure, i.e., that possess internal degrees of freedom) couples to the {\it non-Riemannian} geometric quantities, see the second term on the r.h.s.\ of (\ref{Ptot}) and the first term on the r.h.s.\ of (\ref{DY_pd_prop_eq_1a}). In contrast to this, the {\it orbital} canonical energy-momentum (which is induced by the {\it macroscopic} dynamics of the rotating and deformable body) is only coupled to the purely Riemannian  geometric variables and never couples to the non-Riemannian geometry, see the last terms on the right-hand sides of (\ref{Ptot}) and (\ref{DY_pd_prop_eq_1a}). 

This observation demonstrates that the possible presence of the non-Riemannian geometry (in particular, of torsion and nonmetricity) can only be tested with the help of bodies that are constructed from media with microstructure (spin, dilaton charge and intrinsic shear). This confirms and generalizes the result in \cite{Stoeger:Yasskin:1980}. Test particles, composed from usual matter {\it without} microstructure, are {\it not} affected by the non-Riemannian geometry, and they thus {\it cannot} be used for the detection of the torsion and the nonmetricity. 

These results should be taken into account in the design of future experiments aimed to test the geometric nature of spacetime. Such experiments necessarily have to use microstructured test bodies (a spin-polarized sphere or a polarized beam of elementary particles, e.g.) in order to be able to detect non-Riemannian spacetime features. Technological challenges in this context concern the construction of suitable devices, most importantly, replacing the mechanical gyroscopes with, for example, nuclear magnetic resonance gyroscopes which -- since the 1960's (see \cite{Simpson:1964}, e.g.) -- utilize the spin of nuclei for the purpose of inertial navigation.

\section{Special case: Hayashi-Shirafuji model.}

Our conclusions are very general and apply to all gravitational models that belong to the framework of MAG. The tetrad gravity models studied in \cite{Mao:etal:2006} are special MAG theories, and the measurement of the torsion by means of usual gyroscopes is strictly ruled out for these models: There is no way to detect and/or place limits on the spacetime torsion with the Gravity Probe B mission (see also the relevant analysis in \cite{Flanagan:Rosenthal:2007}).

This point seems to be unclear to and underestimated by the authors of the recent paper \cite{Mao:etal:2006}, who claim that the gravitational model of Hayashi and Shirafuji \cite{Hayashi:Shirafuji:1979} may have special properties that allow for the detection of the torsion with the help of usual gyroscopes. Here we explicitly demonstrate that this claim is unsubstantiated. 

The Hayashi-Shirafuji model is naturally embedded into the MAG scheme as follows (see also \cite{Obukhov:Pereira:2003}). Of the three variables $(h^\alpha_i, \Gamma_{i\beta}{}^\alpha, g_{\alpha\beta})$, the tetrad (coframe) field $h^\alpha_i$ is treated as a translational gauge potential of the gravitational field, whereas the local linear connection $\Gamma_{i\beta}{}^\alpha$ and the metric $g_{\alpha\beta}$ play a secondary role due to the geometrical (teleparallelism) constraints imposed on the spacetime manifold. The torsion $S_{ij}{\,}^\alpha = D_ih^\alpha_j - D_jh^\alpha_i$ is interpreted as the translational gauge field strength. The covariant derivative is defined here as $D_ih^\alpha_j = \partial_ih^\alpha_j + \Gamma_{i\beta}{}^\alpha h^\beta_j$, and the operator $D_i$ acts in a similar covariant way on all {\it tetrad indices} (denoted by Greek letters).

The action of the Hayashi-Shirafuji model $I = \int {\cal L}\,d^4x$ is determined by the Lagrangian density ${\cal L} = \sqrt{-g}\,L$ which is quadratic in torsion,
\begin{equation}\label{LHS}
L = -\,{\frac 14}\left(c_1\,S_{ij}{\,}^\alpha S^{ij}{\,}_\alpha + c_2\,S_i S^i + c_3\,S_{ij}{\,}^\alpha S^i{\,}_\alpha{}^j\right).
\end{equation}
Here $c_1, c_2, c_3$ represent three coupling constants. The torsion trace vector is defined as $S_i:= S_{ij}{\,}^\alpha h^j_\alpha$, and we convert freely the Greek (tetrad) indices into the Latin (coordinate) ones and vice versa by transvection with tetrads. 

As usual, $g := {\rm det}g_{ij}$. In general the tetrad legs are not orthonormal, hence the metric $g_{\alpha\beta} = h^i_\alpha h^j_\beta g_{ij}$ -- which describes the scalar products of the tetrad vectors -- is not a constant matrix but a function of the spacetime coordinates.

The Lagrangian is a function of the three variables, ${\cal L} = {\cal L}(h^\alpha_i, \Gamma_{i\beta}{}^\alpha, g_{\alpha\beta})$, and accordingly we have three variational derivatives that we denote by
\begin{equation}
{\cal E}_\alpha{}^i := {\frac {\delta {\cal L}}{\delta h_i^\alpha}},\quad {\cal C}^{i\beta}{}_\alpha := {\frac {\delta {\cal L}}{\delta \Gamma_{i\beta} {}^\alpha}},\quad {\cal G}^{\alpha\beta} := 2{\frac {\delta {\cal L}} {\delta g_{\alpha\beta}}}.\label{ECG}
\end{equation}
In Appendices A, B, and C we show that for the Hayashi-Shirafuji model (\ref{LHS}) these derivatives satisfy two {\it strong identities} 
\begin{eqnarray}
h^\alpha_k D_i{\cal E}_\alpha{}^i \equiv S_{ki}{}^\alpha\,{\cal E}_\alpha{}^i + R_{ki\beta}{}^\alpha\,{\cal C}^{i\beta}{}_\alpha - {\frac 12}Q_{k\alpha\beta} \,{\cal G}^{\alpha\beta},\label{dE}\\
D_i{\cal C}^{i\beta}{}_\alpha + h_i^\beta\,{\cal E}_\alpha{}^i - {\cal G}^\beta{}_\alpha \equiv 0.\label{dC}
\end{eqnarray}

The total system of the interacting gravitational and matter fields is described by the Lagrangian ${\cal L} + {\cal L}_{\rm mat}$. With the matter sources defined by
\begin{equation}
\widetilde{T}_\alpha{}^i := {\frac {\delta {\cal L_{\rm mat}}}{\delta h_i^\alpha}},\quad\widetilde{\Delta}^\beta{}_\alpha{}^i := {\frac {\delta {\cal L_{\rm mat}}}{\delta \Gamma_{i\beta}{}^\alpha}},\quad \widetilde{t}^{\alpha\beta} := 2{\frac {\delta {\cal L_{\rm mat}}}{\delta g_{\alpha\beta}}},\label{TDt}
\end{equation}
the gravitational field equations then read 
\begin{equation}
{\cal E}_\alpha{}^i + \widetilde{T}_\alpha{}^i =0,\quad {\cal C}^{i\beta}{}_\alpha + \widetilde{\Delta}^\beta{}_\alpha{}^i =0,\quad {\cal G}^{\alpha\beta} + \widetilde{t}^{\alpha\beta} = 0.\label{HSeq}
\end{equation}
Using these equations in the identities (\ref{dE}) and (\ref{dC}), we obtain the two conservation laws (\ref{em_conservation_ala_DY_1}) and (\ref{em_conservation_ala_DY_2}), after taking into account that $h^k_\alpha D_ih^\alpha_j = \Gamma_{ij}{}^k$ and $Q_{i\alpha\beta}= -\,2N_{i\alpha\beta}$. 

In other words, contrary to the claim of \cite{Mao:etal:2006}, the Hayashi-Shirafuji gravitational theory does not have any special properties. The matter source currents in the Hayashi-Shirafuji model, just like in all other MAG models, satisfy the conservation laws (\ref{em_conservation_ala_DY_1}) and (\ref{em_conservation_ala_DY_2}) which were the starting point for our multipole-moment analysis of the propagation equations. 

\section{Conclusions.}

In \cite{Mao:etal:2006}, the propagation equations are not derived from first principles but are arbitrarily postulated instead. As we have shown explicitly in the previous sections, such an ad-hoc procedure is not compatible with the equations of motion as derived with the help of a multipole method. Like in Einstein's general relativity theory, in the gauge-theoretic models that belong to the MAG scheme the propagation equations need not (and cannot) be postulated separately. They follow directly from the conservation laws of the energy-momentum (for structureless matter) and from the conservation law of the hypermomentum (for matter with microstructure). 

We consistently derived the propagation equations from (\ref{em_conservation_ala_DY_1}) and (\ref{em_conservation_ala_DY_2}) using the systematic multipole expansion method. The conservation laws (\ref{em_conservation_ala_DY_1}) and (\ref{em_conservation_ala_DY_2}) hold for all MAG models, and for the Hayashi-Shirafuji tetrad gravity, in particular. Hence, all mathematical derivations and physical conclusions are valid for the latter model as well. Our analysis shows that the non-Riemannian spacetime geometry can be detected only with the help of matter with microstructure. We thus confirm and generalize the earlier observations of Yasskin and Stoeger \cite{Stoeger:Yasskin:1980}. 

In connection with our results, it seems interesting to reanalyze the axiomatic schemes of Marzke-Wheeler \cite{Marzke:Wheeler:1964:I} and Ehlers-Pirani-Schild \cite{Ehlers:Pirani:Schild:1972,Perlick:1987} in which the geometrical structure of the spacetime is operationally deduced from assumptions about the behavior of {\it primitive measuring devices} (test bodies and light) in the gravitational field. Such axiomatics leads to a Weyl geometry that is characterized by vanishing torsion but has a nontrivial nonmetricity $Q_{ijk} = Q_i g_{jk}$ (with the so-called Weyl covector $Q_i$). How can this fact be understood in the light of the results obtained here? In order to find an answer to this question, one needs to inspect more carefully the definition of the primitive devices. In \cite{Ehlers:Pirani:Schild:1972}, see page 76, they are described very generally as a ``$\dots$ class of test particles (neutral, spherically symmetrical ones) $\dots$". It thus appears that despite the absence of an explicit notice, the axiomatics of Ehlers-Pirani-Schild tacitly assumes the use of test bodies with a special type of microstructure, namely, of the {\it dilationally deformable} bodies. The dilation (isotropic expansion/contraction without distorsion) is clearly compatible with the spherical symmetry of the particles. On the other hand, in the gauge approach of MAG, the generator of dilations is associated precisely with the nonmetricity of the Weylian type. Hence, this would rather naturally explain why the axiomatic scheme leads to the non-Riemannian geometry of Weyl. Of course, a more detailed analysis is needed in order to check whether this holds for all possible devices described in \cite{Ehlers:Pirani:Schild:1972}. 

There remain several theoretical questions that need to be addressed in the context of the multipole approximation of the equations of motion in metric-affine gravity. In particular, theoretical challenges concern: the specification of a world line of the body, the invariant definition of multipole moments, the identification of objects which have well-defined classical limits, the control of higher orders in the multipole expansion, and the role of supplementary or constitutive relations. Regarding the last point, previous analyses \cite{Frenkel:1926,Corinaldesi:Papapetrou:1951,Pirani:1956,Tulczyjew:1959} in metric theories of gravitation have shown that one needs -- already at the dipole level -- to impose a supplementary condition in order to obtain a closed set of propagation equations. Since -- in contrast to metric theories as well as theories on a Riemann-Cartan background -- the spectrum of possible supplementary conditions in MAG is greatly enhanced, we hope to be able to present a systematic analysis of different conditions in a future work. 

The works of Babourova and Frolov \cite{Babourova:Frolov:1998,Babourova:Frolov:1998:2} make a preliminary step in this direction by analyzing the supplementary conditions in the model of matter with a particular constitutive structure -- the ideal fluid with microstructure. Their conclusions agree completely with our results and confirm the impossibility to detect the non-Riemannian geometry by means of the ordinary matter.

\section{Acknowledgments.}
The authors are grateful to F.W.\ Hehl (Univ.\ Cologne) for stimulating discussions and constructive criticism. YNO was supported by the Deutsche Forschungsgemeinschaft (Bonn) with the grant HE 528/21-1. DP acknowledges the support by {\O}.\ Elgar{\o}y (Univ.\ Oslo) and the Research Council of Norway under the project 162830 as well as the support by the Deutsche Forschungsgemeinschaft (Bonn) under the grant SFB/TR 7.

\section*{Appendix A: Basic geometrical identities.}

We recall that the curvature arises (can be defined) from the commutator of the covariant derivatives, $(D_iD_j - D_jD_i)V^\alpha = R_{ij\beta}{}^\alpha V^\beta$. Since the torsion is $S_{ij}{\,}^\alpha = D_ih^\alpha_j - D_jh^\alpha_i$, and the nonmetricity is $Q_{i\alpha\beta} = -D_ig_{\alpha\beta}$, one can straightforwardly verify the Bianchi identities:
\begin{eqnarray}
D_iR_{jk\beta}{}^\alpha + D_jR_{ki\beta}{}^\alpha + D_kR_{ij\beta}{}^\alpha \equiv 0,\label{B1}\\
D_iS_{jk}{}^\alpha + D_jS_{ki}{}^\alpha + D_kS_{ij}{}^\alpha \equiv R_{ijk}{}^\alpha + R_{jki}{}^\alpha + R_{kij}{}^\alpha,\label{B2}\\
D_iQ_{j\alpha\beta} - D_jQ_{i\alpha\beta} \equiv 2R_{ij(\alpha\beta)}.{\ }\label{B0}
\end{eqnarray}

\section*{Appendix B: Algebraic identities.}

We now turn to the specific tetrad theory which was discussed in \cite{Mao:etal:2006}. 

The Lagrangian density ${\cal L}$ of the Hayashi-Shirafuji model (\ref{LHS}) is a function of the torsion, metric, and the tetrad. The partial derivatives with respect to these arguments are easily computed:
\begin{eqnarray}
H^{ij}{}_\alpha &=& -2\,{\frac {\partial {\cal L}}{\partial S_{ij}{\,}^\alpha}}\label{Hij0}\\ 
&=& \sqrt{-g}\left( c_1 S^{ij}{}_\alpha + c_2 S^{[i}h^{j]}_\alpha + c_3 S^{[i}{}_\alpha{}^{j]}\right),\label{Hij}\\
{\frac {\partial {\cal L}}{\partial g_{\alpha\beta}}} &=& {\frac {\sqrt{-g}}2} \Big[ c_1 (S^{\alpha k}{}_\gamma S^\beta{}_k{}^\gamma - {\frac 12}\,S_{kl}{}^\alpha S^{kl\beta})\nonumber \\
&& +\,{\frac {c_2}2}\,S^\alpha S^\beta + {\frac {c_3}2}\,S^\alpha{}_\gamma{}^\delta S^\beta{}_\delta{}^\gamma + L\,g^{\alpha\beta}\Big],\label{dLg}\\
{\frac {\partial {\cal L}}{\partial h_i^\alpha}} &=& {\sqrt{-g}}\Big[c_1 S_{\alpha k}{}^\gamma S^{ik}{}_\gamma + {\frac {c_2}2}\,( S_\alpha S^i + S_{k\alpha}{}^i S^k ) \nonumber \\
&& +\,{\frac {c_3}2}\,( S^i{}_\gamma{}^k S_{\alpha k}{}^\gamma + S^k{}_\gamma{}^i S_{k\alpha}{}^\gamma ) + L\,h_\alpha^i \Big].\label{dLh} 
\end{eqnarray}
The direct check shows that these three quantities satisfy the two following {\it algebraic identities}:
\begin{eqnarray}
h_k^\alpha\,{\frac {\partial {\cal L}}{\partial h_i^\alpha}} - {\cal L}\,\delta_k^i - H^{ij}{}_\alpha S_{kj}{}^\alpha &\equiv& 0,\label{id1}\\
2\,g_{\alpha\gamma}\,{\frac {\partial {\cal L}}{\partial g_{\beta\gamma}}}- h_i^\beta\,{\frac {\partial {\cal L}}{\partial h_i^\alpha}} + {\frac 12}\,H^{ij}{}_\alpha S_{ij}{}^\beta &\equiv& 0.\label{id2}
\end{eqnarray}
It is worthwhile to stress that these relations hold {\it identically}, without taking into account the field equations. 

The {\it variational derivatives} (\ref{ECG}) with respect to the gravitational potentials read explicitly:
\begin{eqnarray}
{\cal E}_\alpha{}^i &=&{\frac {\partial {\cal L}}{\partial h_i^\alpha}} - D_jH^{ij}{}_\alpha,\label{E}\\
{\cal C}^{i\beta}{}_\alpha &=& {\frac {\partial {\cal L}}{\partial S_{kl}{}^\gamma}}\,{\frac {\partial S_{kl}{}^\gamma}{\partial \Gamma_{i\beta}{}^\alpha}} = -\,H^{i\beta}{}_\alpha,\label{C}\\
{\cal G}^{\alpha\beta} &=& 2{\frac {\partial {\cal L}}{\partial g_{\alpha\beta}}}.\label{G}
\end{eqnarray}

\section*{Appendix C: Differential identities.}

Using the chain rule for ${\cal L} = {\cal L}(h_i^\alpha, g_{\alpha\beta}, S_{ij}{}^\alpha)$, we have
\begin{equation}
\partial_k{\cal L} = {\frac {\partial {\cal L}}{\partial h_i^\alpha}}\,\partial_k h_i^\alpha + {\frac {\partial {\cal L}}{\partial g_{\alpha\beta}}}\,\partial_k g_{\alpha\beta} + {\frac {\partial {\cal L}}{\partial S_{ij}{}^\alpha}}\,\partial_k S_{ij}{}^\alpha.\label{dL}
\end{equation}
Furthermore, by differentiating the identity (\ref{id1}) (apply the operator $\partial_i$), we find
\begin{equation}
\partial_i\left(h_k^\alpha\,{\frac {\partial {\cal L}}{\partial h_i^\alpha}}\right) \equiv \partial_k{\cal L} + (D_iH^{ij}{}_\alpha) S_{kj}{\,}^\alpha + H^{ij}{}_\alpha D_iS_{kj}{\,}^\alpha.\label{id1d}
\end{equation}
Contracting (\ref{id2}) with $\Gamma_{k\beta}{}^\alpha$ yields
\begin{equation}
2\,{\frac {\partial {\cal L}}{\partial g_{\alpha\beta}}}\,\Gamma_{k(\alpha\beta)}- {\frac {\partial {\cal L}}{\partial h_i^\alpha}}\Gamma_{k\beta}{}^\alpha h_i^\beta\, + {\frac 12}\,H^{ij}{}_\alpha S_{ij}{\,}^\beta\Gamma_{k\beta}{}^\alpha \equiv 0.\label{id2G}
\end{equation}

Now we are ready to derive the main differential identities. The covariant divergence of (\ref{E}) reads:
\begin{equation}
D_i{\cal E}_\alpha{}^i = D_i{\frac {\partial {\cal L}}{\partial h_i^\alpha}} - D_iD_jH^{ij}{}_\alpha.\label{dE1}
\end{equation}
Taking into account the skew symmetry $H^{ij}{}_\alpha = -\,H^{ji}{}_\alpha$, and the fact that the commutator of the covariant derivatives, $D_iD_j - D_jD_i$, produces the curvature in the last term, after contracting (\ref{dE1}) with $h^\alpha_k$, we find
\begin{equation}
h^\alpha_k D_i{\cal E}_\alpha{}^i = \partial_i\left(h_k^\alpha\,{\frac{\partial {\cal L}}{\partial h_i^\alpha}}\right) - {\frac {\partial {\cal L}}{\partial h_i^\alpha}}\,D_ih^\alpha_k + {\frac 12}\,R_{ijk}{}^\alpha H^{ij}{}_\alpha.\label{dE2}
\end{equation}
The first term was transformed with the help of the Leibniz rule. For the second term on the right-hand side, we note that $D_ih^\alpha_k = D_ih^\alpha_k - D_kh^\alpha_i + D_kh^\alpha_i = S_{ik}{}^\alpha + D_kh^\alpha_i$. As for the last term on the right-hand side of (\ref{dE2}), we transform it using the Bianchi identity (\ref{B2}) into
\begin{eqnarray}
{\frac 12}\,R_{ijk}{}^\alpha H^{ij}{}_\alpha &=& -\,R_{ki\beta}{}^\alpha H^{i\beta}{}_\alpha - H^{ij}{}_\alpha D_iS_{kj}{\,}^\alpha \nonumber\\
&& +\,{\frac 12}\,H^{ij}{}_\alpha D_k S_{ij}{\,}^\alpha.\label{dE3}
\end{eqnarray}
Taking this into account, and substituting (\ref{id1d}), we recast (\ref{dE2}) into
\begin{eqnarray}
h^\alpha_k D_i{\cal E}_\alpha{}^i &\equiv& \partial_k{\cal L} - {\frac {\partial{\cal L}}{\partial h_i^\alpha}}\,D_kh_i^\alpha + {\frac 12}\,H^{ij}{}_\alpha D_k S_{ij}{\,}^\alpha \nonumber\\ \label{dE4}
&-& R_{ki\beta}{}^\alpha H^{i\beta}{}_\alpha + {\frac {\partial {\cal L}}{\partial h_i^\alpha}}\,S_{ki}{}^\alpha - (D_jH^{ij}{}_\alpha) S_{ki}{\,}^\alpha.
\end{eqnarray}
Furthermore, we have 
\begin{eqnarray}
&&-\,{\frac {\partial {\cal L}}{\partial h_i^\alpha}}\,D_kh_i^\alpha + {\frac 12}\,H^{ij}{}_\alpha D_k S_{ij}{\,}^\alpha = - {\frac {\partial {\cal L}}{\partial h_i^\alpha}}\,\partial_kh_i^\alpha \nonumber\\
&& +\,{\frac 12}\,H^{ij}{}_\alpha \partial_k S_{ij}{\,}^\alpha - {\frac {\partial {\cal L}}{\partial h_i^\alpha}}\Gamma_{k\beta}{}^\alpha h_i^\beta\, + {\frac 12}\,H^{ij}{}_\alpha S_{ij}{\,}^\beta\Gamma_{k\beta}{}^\alpha \nonumber\\
&&= -\,{\frac {\partial {\cal L}}{\partial h_i^\alpha}}\,\partial_kh_i^\alpha + {\frac 12}\,H^{ij}{}_\alpha \partial_k S_{ij}{\,}^\alpha - 2\,{\frac {\partial {\cal L}}{\partial g_{\alpha\beta}}}\,\Gamma_{k(\alpha\beta)}.\label{dE5}
\end{eqnarray}
Here we used the identity (\ref{id2G}). Recalling the definition of the nonmetricity, $Q_{k\alpha\beta} = - D_kg_{\alpha\beta} = - \partial_kg_{\alpha\beta} + 2\Gamma_{k(\alpha\beta)}$, and the definition of the field momentum (\ref{Hij0}), we get
\begin{eqnarray}
&&-\,{\frac {\partial {\cal L}}{\partial h_i^\alpha}}\,D_kh_i^\alpha + {\frac 12} \,H^{ij}{}_\alpha D_k S_{ij}{\,}^\alpha = -\,{\frac {\partial{\cal L}}{\partial g_{\alpha\beta}}}\,Q_{k\alpha\beta} \nonumber\\ \label{dE6}
&& - {\frac {\partial {\cal L}}{\partial h_i^\alpha}}\,\partial_kh_i^\alpha - {\frac {\partial{\cal L}}{\partial S_{ij}{}^\alpha}}\,\partial_k S_{ij}{\,}^\alpha -\,{\frac {\partial {\cal L}}{\partial g_{\alpha\beta}}}\,\partial_kg_{\alpha\beta}.
\end{eqnarray}
Substituting this into (\ref{dE4}) and taking into account (\ref{dL}), we finally arrive at
\begin{eqnarray}
h^\alpha_k D_i{\cal E}_\alpha{}^i &\equiv& S_{ki}{}^\alpha\left({\frac {\partial {\cal L}}{\partial h_i^\alpha}} - D_jH^{ij}{}_\alpha\right)\nonumber\\
&& -\,R_{ki\beta}{}^\alpha H^{i\beta}{}_\alpha - {\frac {\partial {\cal L}} {\partial g_{\alpha\beta}}}\,Q_{k\alpha\beta}.\label{dE7}
\end{eqnarray}
Recalling the definition of the variational derivatives (\ref{E})-(\ref{G}), we finally recast this identity into (\ref{dE}).

The second differential identity is derived more straightforwardly. We take (\ref{E}) and contract it with $h_i^\beta$:
\begin{eqnarray}
h_i^\beta\,{\cal E}_\alpha{}^i &=& h_i^\beta\,{\frac {\partial{\cal L}}{\partial h_i^\alpha}} - h_i^\beta\,D_jH^{ij}{}_\alpha \nonumber\\
&=& h_i^\beta\,{\frac {\partial {\cal L}}{\partial h_i^\alpha}} - D_jH^{\beta j}{}_\alpha + (D_jh_i^\beta)\,H^{ij}{}_\alpha \nonumber\\
&=& h_i^\beta\,{\frac {\partial {\cal L}}{\partial h_i^\alpha}} + D_iH^{i\beta}{}_\alpha - {\frac 12}S_{ij}{}^\beta\,H^{ij}{}_\alpha \nonumber\\ 
&=& D_iH^{i\beta}{}_\alpha + 2\,g_{\alpha\gamma}\,{\frac {\partial {\cal L}}{\partial g_{\beta\gamma}}}.\label{dC1}
\end{eqnarray}
In the last equality we used (\ref{id2}). With the definitions (\ref{C}) and (\ref{G}), we thus prove the identity (\ref{dC}). 

\bibliographystyle{unsrt}
\bibliography{micro_bibliography}

\begin{thebibliography}{10}

\bibitem{Mathisson:1931:3}
M.~Mathisson.
\newblock {Bewegungsproblem der Feldphysik und Elektronenkonstanten}.
\newblock {\em Z. Phys.}, 69:389, 1931.

\bibitem{Mathisson:1931:1}
M.~Mathisson.
\newblock {Die Beharrungsgesetze der allgemeinen Relativit\"{a}tstheorie}.
\newblock {\em Z. Phys.}, 67:270, 1931.

\bibitem{Mathisson:1931:2}
M.~Mathisson.
\newblock {Die Mechanik des Materieteilchens in der allgemeinen
  Relativit\"{a}tstheorie}.
\newblock {\em Z. Phys.}, 67:826, 1931.

\bibitem{Robertson:1937}
H.~P. {Robertson}.
\newblock {Test corpuscles in general relativity}.
\newblock {\em Proc. Edn. Math. Soc.}, 5:63, 1937.

\bibitem{Fock:1939}
V.~A. {Fock}.
\newblock {Sur le mouvement des masses finies d'apr\`{e}s la th\'{e}orie de
  gravitation einsteinienne}.
\newblock {\em J. Phys. (Moscow)}, 1:81, 1939.

\bibitem{Mathisson:1937}
M.~Mathisson.
\newblock {Neue Mechanik materieller Systeme}.
\newblock {\em Acta Phys. Pol.}, 6:163, 1937.

\bibitem{Papapetrou:1940:1}
A.~Papapetrou.
\newblock {Gravitationswirkungen zwischen Pol-Dipol Teilchen}.
\newblock {\em Z. Phys.}, 116:298, 1940.

\bibitem{InfeldSchild:1949}
L.~Infeld and A.~Schild.
\newblock {On the motion of test particles in General Relativity}.
\newblock {\em Rev. Mod. Phys.}, 21:408, 1949.

\bibitem{Papapetrou:1951:3}
A.~Papapetrou.
\newblock {Spinning test-particles in General Relativity. I}.
\newblock {\em Proc. Royal Soc. London Ser. A. Math. and Phys. Sci.}, 209:248,
  1951.

\bibitem{Papapetrou:1951}
A.~Papapetrou.
\newblock {Equations of motion in General Relativity}.
\newblock {\em Proc. Phys. Soc. A}, 64:57, 1951.

\bibitem{Tulczyjew:1959}
W.~Tulczyjew.
\newblock {Motion of multipole particles in general relativity theory}.
\newblock {\em Acta Phys. Pol.}, 18:393, 1959.

\bibitem{HavasGoldberg:1962}
P.~{Havas} and J.~N. {Goldberg}.
\newblock {Lorentz-invariant equations of motion of point masses in the general
  theory of relativity}.
\newblock {\em Phys. Rev.}, 128:398, 1962.

\bibitem{Mao:etal:2006}
Y.~{Mao}, M.~{Tegmark}, A.~{Guth}, and S.~{Cabi}.
\newblock {Constraining torsion with Gravity Probe B}.
\newblock {\em Phys. Rev. D}, 76:104029, 2007.

\bibitem{Hehl:1995}
F.~W. {Hehl}, J.~D. {McCrea}, E.~W. {Mielke}, and Y.~{Ne'eman}.
\newblock {Metric-affine gauge theory of gravity: Field equations, Noether
  identities, world spinors, and breaking of dilation invariance}.
\newblock {\em Phys. Rep.}, 258:1, 1995.

\bibitem{Hehl:1971}
F.~W. {Hehl}.
\newblock {How does one measure torsion of space-time?}
\newblock {\em Phys. Lett. A}, 36:225, 1971.

\bibitem{Trautman:1972}
A.~Trautman.
\newblock {On the Einstein-Cartan equations III}.
\newblock {\em Bull. Acad. Pol. Sci.}, 20:895, 1972.

\bibitem{Stoeger:Yasskin:1979}
W.~R. {Stoeger} and P.~B. {Yasskin}.
\newblock {Can a macroscopic gyroscope feel torsion?}
\newblock {\em Gen. Relativ. Gravit.}, 11:427, 1979.

\bibitem{Stoeger:Yasskin:1980}
P.~B. {Yasskin} and W.~R. {Stoeger}.
\newblock {Propagation equations for test bodies with spin and rotation in
  theories of gravity with torsion}.
\newblock {\em Phys. Rev. D}, 21:2081, 1980.

\bibitem{Nomura:Shirafuji:Hayashi:1991}
K.~{Nomura}, T.~{Shirafuji}, and K.~{Hayashi}.
\newblock {Spinning test particles in spacetime with torsion}.
\newblock {\em Prog. Theo. Phys.}, 86:1239, 1991.

\bibitem{Audretsch:1981:1}
J.~{Audretsch}.
\newblock {Dirac electron in space-times with torsion: Spinor propagation, spin
  precession, and nongeodesic orbits}.
\newblock {\em Phys. Rev. D}, 24:1470, 1981.

\bibitem{Bragov:etal:1991:1}
V.~G. {Bagrov}, V.~V. {Belov}, A.~Y. {Trifonov}, and A.~Y. {Yevseyevich}.
\newblock {The complex WKB-Maslov method for the Dirac equation in a torsion
  field. 1. Construction of trajectory-coherent states and the equation for
  spin}.
\newblock {\em Class. Quantum Grav.}, 8:1349, 1991.

\bibitem{Bragov:etal:1991:2}
V.~G. {Bagrov}, V.~V. {Belov}, A.~Y. {Trifonov}, and A.~Y. {Yevseyevich}.
\newblock {Quasi-classical trajectory-coherent approximation for the Dirac
  equation with an external electromagnetic field in Riemann-Cartan space. 2.
  Construction of $TCS$ and equation of motion for spin}.
\newblock {\em Class. Quantum Grav.}, 8:1833, 1991.

\bibitem{Nomura:Shirafuji:Hayashi:1992}
K.~{Nomura}, T.~{Shirafuji}, and K.~{Hayashi}.
\newblock {Semiclassical particles with arbitrary spin in the Riemann-Cartan
  space-time}.
\newblock {\em Prog. Theo. Phys.}, 87:1275, 1992.

\bibitem{Shapiro:2002}
I.~L. {Shapiro}.
\newblock {Physical aspects of the space-time torsion}.
\newblock {\em Phys. Rep.}, 357:113, 2002.

\bibitem{Cosserat:1909}
E.~Cosserat and F.~Cosserat.
\newblock {\em {Th\'eorie des corps d\'eformables}}.
\newblock Hermann, Paris, 1909.

\bibitem{Weyssenhoff:Raabe:1947}
J.~Weyssenhoff and A.~Raabe.
\newblock {Relativistic dynamics of spin-fluids and spin-particles}.
\newblock {\em Acta Phys. Pol.}, 9:7, 1947.

\bibitem{Kroener:1958}
E.~{Kr\"oner}.
\newblock {Kontinuumstheorie der Versetzungen und Eigenspannungen}.
\newblock {\em Ergebnisse der angewandten Mathematik, Eds. L. Collatz, F.
  L\"osch, Springer, Berlin}, 5, 1958.

\bibitem{Truesdell:Toupin:1960}
C.~{Truesdell} and R.~A. {Toupin}.
\newblock {The classical field theories}.
\newblock {\em Handbuch der Physik, Ed. S. Fl{\"u}gge, Springer, Berlin},
  III/1:226, 1960.

\bibitem{Mindlin:1964}
R.~D. {Mindlin}.
\newblock {Micro-structure in linear elasticity}.
\newblock {\em Arch. Rational Mech. Anal.}, 16:51, 1964.

\bibitem{Capriz:1989}
G.~Capriz.
\newblock {\em {Continua with microstructure}}.
\newblock Springer Tracts in Natural Philosophy, Springer, Berlin, 1989.

\bibitem{Gronwald:Hehl:1996}
F.~{Gronwald} and F.~W. {Hehl}.
\newblock {On the gauge aspects of gravity}.
\newblock {\em Proc. Int. School of Cosmology and Gravitation: 14th Course,
  Erice, Italy, Eds. P.G. Bergmann et al. (World Scientific, Singapore)}, page
  148, 1996.

\bibitem{Obukhov:Rubilar:2006}
Y.~N. {Obukhov} and G.~F. {Rubilar}.
\newblock {Invariant conserved currents in gravity theories with local Lorentz
  and diffeomorphism symmetry}.
\newblock {\em Phys. Rev. D}, 74:064002, 2006.

\bibitem{Simpson:1964}
J.~H. {Simpson}.
\newblock {Nuclear gyroscopes}.
\newblock {\em Astron. Aeron.}, 2:42, 1964.

\bibitem{Flanagan:Rosenthal:2007}
{\'E}.~{\'E}. {Flanagan} and E.~{Rosenthal}.
\newblock {Can Gravity Probe B usefully constrain torsion gravity theories?}
\newblock {\em Phys. Rev. D}, 75:124016, 2007.

\bibitem{Hayashi:Shirafuji:1979}
K.~{Hayashi} and T.~{Shirafuji}.
\newblock {New general relativity}.
\newblock {\em Phys. Rev. D}, 19:3524, 1979.

\bibitem{Obukhov:Pereira:2003}
Y.~N. {Obukhov} and J.~G. {Pereira}.
\newblock {Metric-affine approach to teleparallel gravity}.
\newblock {\em Phys. Rev. D}, 67:044016, 2003.

\bibitem{Marzke:Wheeler:1964:I}
R.~F. {Marzke} and J.~A. {Wheeler}.
\newblock {Gravitation as geometry. I: The geometry of space-time and the
  geometrodynamical standard meter}.
\newblock {\em Gravitation and Relativity, Eds. H.-Y. Chiu and W.F. Hoffmann,
  W.A. Benjamin, New York}, page~40, 1964.

\bibitem{Ehlers:Pirani:Schild:1972}
J.~{Ehlers}, F.~A.~E. {Pirani}, and A.~{Schild}.
\newblock {The geometry of free fall and light propagation}.
\newblock {\em General Relativity, Ed. L. O'Raifeartaigh, Oxford Univ. Press,
  New York}, page~63, 1972.

\bibitem{Perlick:1987}
V.~{Perlick}.
\newblock {Characterization of standard clocks by means of light rays and
  freely falling particles}.
\newblock {\em Gen. Rel. Grav.}, 19:1059, 1987.

\bibitem{Frenkel:1926}
J.~Frenkel.
\newblock {Die Elektrodynamik des rotierenden Elektrons}.
\newblock {\em Z. Phys.}, 37:243, 1926.

\bibitem{Corinaldesi:Papapetrou:1951}
E.~Corinaldesi and A.~Papapetrou.
\newblock {Spinning test-particles in General Relativity. II}.
\newblock {\em Proc. Royal Soc. London Ser. A. Math. and Phys. Sci.}, 209:259,
  1951.

\bibitem{Pirani:1956}
F.~A.~E. Pirani.
\newblock {On the physical significance of the Riemann tensor}.
\newblock {\em Acta Phys. Pol.}, 15:389, 1956.

\bibitem{Babourova:Frolov:1998}
O.~V. {Babourova} and B.~N. {Frolov}.
\newblock {Perfect hypermomentum fluid: Variational theory and equations of
  motion}.
\newblock {\em Int. J. Mod. Phys. A}, 13:5391, 1998.

\bibitem{Babourova:Frolov:1998:2}
O.~V. {Babourova} and B.~N. {Frolov}.
\newblock {Perfect fluid and test particle with spin and dilatonic charge in a
  Weyl-Cartan space}.
\newblock {\em Mod. Phys. Lett. A}, 13:7, 1998.

\end{thebibliography}

\end{document}